\documentclass[letterpaper, prl, showpacs]{revtex4}
\usepackage[T1]{fontenc}
\usepackage{graphics, graphicx, amsmath, pifont, color, setspace}

\begin{document}
\title{Single-photon cooling to the limit of trap dynamics: Maxwell's Demon near maximum efficiency}
\author{S. Travis Bannerman}
\author{Gabriel N. Price}
\author{Kirsten Viering}
\author{Mark G. Raizen}
\email{raizen@physics.utexas.edu}
\affiliation{Center for Nonlinear Dynamics and Department of Physics, The University of Texas at Austin, Austin, Texas 78712, USA}

\begin{abstract}
We demonstrate a general and efficient informational cooling technique for atoms which is an experimental realization of a one-dimensional Maxwell's Demon. The technique transfers atoms from a magnetic trap into an optical trap via a single spontaneous Raman transition which is discriminatively driven near each atom's classical turning point. In this way, nearly all of the atomic ensemble's kinetic energy in one dimension is removed. We develop a simple analytical model to predict the efficiency of transfer between the traps and provide evidence that the performance is limited only by particle dynamics in the magnetic trap. Transfer efficiencies up to $2.2\%$ are reported. We show that efficiency can be traded for phase-space compression, and we report compression up to a factor of 350. Our results represent a 15-fold improvement over our previous demonstration of the cooling technique.
\end{abstract}


\maketitle

\section{Introduction}

The intrinsic connection between information and thermodynamic entropy was first recognized by Leo Szilard in a landmark paper in 1929 \cite{Szilard} and has since become a cornerstone of modern information science \cite{Shannon, Jaynes, Bennett, Scully}.  Szilard introduced this concept to resolve the apparent violation of the second law of thermodynamics in a thought experiment known as Maxwell's Demon \cite{Maxwell}.  A key prediction was that information can be used to reduce the entropy of a gas of particles.  Indeed, measurement and feedback is the basis for stochastic cooling in accelerator rings \cite{Moehl,Meer}.  However, the available information radiated by the charged particles in the ring is enormous compared with the tiny fraction of information that is actually collected and used for cooling.\par

Recently we proposed the concept of a one-way wall for atoms and molecules and showed how it can be used for cooling \cite{Raizen, Dudarev, Price}.  In parallel, an atom diode operating in a similar fashion was independently proposed without application to cooling \cite{Ruschhaupt}.  Such a one-way wall was directly demonstrated in a proof-of-principle experiment \cite{Thorn}.  Our group used these principles to accumulate atoms from a magnetic trap into an optical trap, and we reported cooling and phase-space compression \cite{Price2}.  We call this method ``single-photon cooling'' because each atom scatters only one photon on average for a nearly complete reduction of kinetic energy in one dimension.\par 

\begin{figure}
\begin{center}
\includegraphics[width=0.75\textwidth]{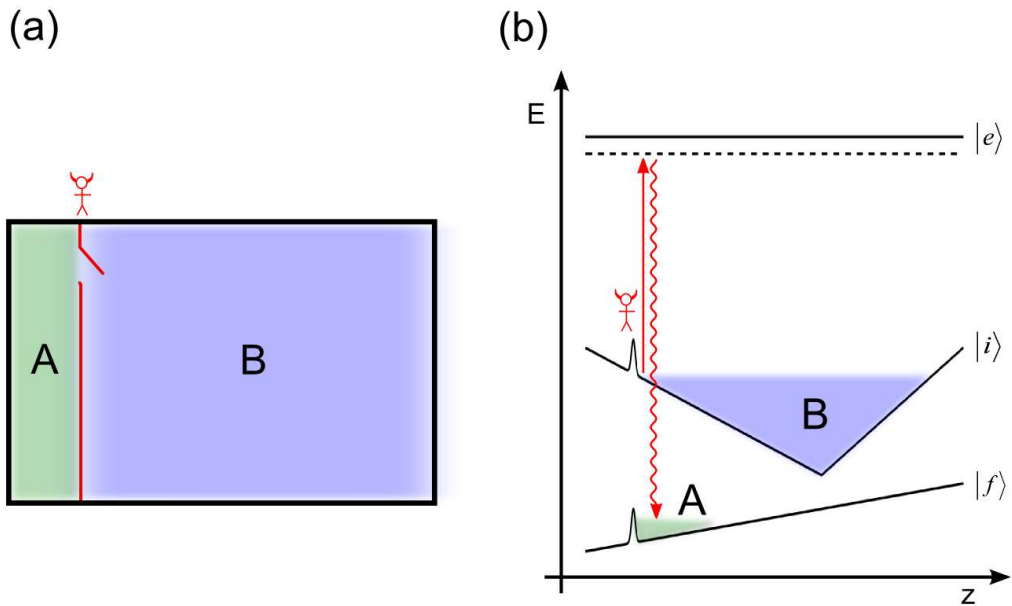}
\end{center}
\caption{\label{fig1}(a) Maxwell's Demon operates a trapdoor between two compartments \textbf{A} and \textbf{B}. By allowing one-way passage of atoms from \textbf{B} to \textbf{A}, the atoms are compressed without expenditure of work. (b) Schematic of single-photon cooling in a three-level system. Magnetically trapped atoms in state $|i\rangle$ are optically pumped via excited state $|e\rangle$ into a non-magnetic state $|f\rangle$ near their classical turning points. Potentials are drawn (not to scale) as a function of the vertical coordinate \textit{z}, with gravity pointing to the left.}
\end{figure}

The operation of a one-way wall for cooling atoms can be understood as a straightforward realization of Maxwell's Demon. 
In the traditional picture, Figure \ref{fig1}(a), the Demon operates a trapdoor between two compartments of atoms, \textbf{A} and \textbf{B}. Without expending any work, the Demon may lower the entropy of the entire system by observing the atoms and allowing the hottest atoms to pass from \textbf{A} to \textbf{B}, and the coldest from \textbf{B} to \textbf{A}. Similarly, if all of the atoms were initially in \textbf{B}, the Demon could lower the entropy of the system by allowing one-way passage from \textbf{B} to \textbf{A}, the smaller compartment. It is clear that the Demon must measure each atom's position to effectively operate the trapdoor, and it is in this nuance that the second law is saved. The informational entropy associated with these measurements compensates for the reduction of thermodynamic entropy.
\par

Our implementation of single-photon cooling is completely analogous to this Demon. A schematic is shown in Figure \ref{fig1}(b). Consider a non-interacting ensemble initially in a low-field-seeking magnetic state $|i\rangle$. The one-dimensional magnetic central potential holding these atoms constitutes \textbf{B}. Atoms are irreversibly transferred by our Demon to \textbf{A}, a gravito-optical trap \cite{Chu}, which is located to the left of (below) \textbf{B}. The Demon in this case is simply a focused pump beam which transfers the atoms from $|i\rangle$ to $|f\rangle$ through a spontaneous Raman transition. This beam is positioned near the classical turning points of the ensemble's most energetic atoms, and the magnetic potential is slowly ramped off. This ensures that each atom is pumped near the turning point of its trajectory, where its kinetic energy vanishes. The Demon thus discriminates the slowest atoms from the rest of the ensemble and releases this informational entropy in the form of a single photon scattered from the pump beam. Should the atom decay to a final state $|f\rangle$ with weaker or opposite magnetic coupling, the potential landscape is altered and a trapped state is produced in the gravito-optical trap. The net result after the pump beam has encountered the entire ensemble is both a reduction in temperature and an increase in density at the ``cost'' of a single photon recoil per atom.\par

One of the key questions regarding the cooling process pertains to efficiency.  One aspect is the efficiency of information entropy used to cool.  We showed, in a conceptual paper, that single-photon cooling is \textit{maximally efficient} in the sense that the entropy increase of the radiation field as each photon is scattered is equal to the entropy reduction of the atoms as they are captured \cite{Ruschhaupt2}.  In this article, we focus on a more utilitarian aspect of efficiency: the fraction of atoms cooled and transferred from the magnetic trap into the optical trap.\par

\section{Experimental Implementation}

We have implemented this general method of informational cooling for  $^{87}$Rb in a three-dimensional quadrupole magnetic trap. The trap is initially populated with atoms in the $5^2S_{1/2}(F=2$) hyperfine manifold, with approximately $70\%$ in the $|F=2, m_F=2\rangle$ state and the remaining in the $|F=2, m_F=1\rangle$ state. We experimentally vary the number $N_B$ and the temperature $T_B$ of atoms in the magnetic trap, but typical values are $N_B \approx 5\times 10^7$ atoms and $T_B\approx 40\:\mu\text{K}$.\par
 
\begin{figure}
\begin{center}
\includegraphics[width=0.45\textwidth]{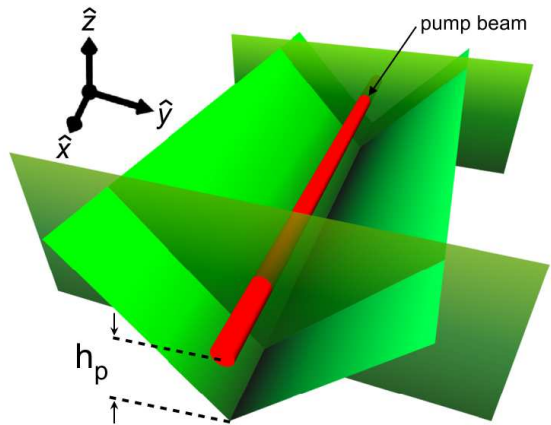}
\end{center}
\caption{\label{fig1c}Configuration of our ``Demon'' and gravito-optical trap.  Two Gaussian laser sheets in a ``V'' shape are orthogonally intersected by two parallel sheets propagating along the $y$ axis. With gravity along the \textit{z} axis, this ``trough'' creates a trapping potential in all three dimensions.  Additionally, a Raman pump beam propagates along the $x$ axis at a height $h_p$ above the vertex of the trough.  The trough and pump beam are positioned below a cloud of magnetically trapped atoms.}
\end{figure}

Figure \ref{fig1c} illustrates the configuration of our ``Demon'' and gravito-optical trap. A pump beam, detuned $35\:\text{MHz}$ below the $5^2S_{1/2}(F=2)\rightarrow 5^2P_{3/2}(F'=1)$ transition, is tightly focused inside an ``optical trough.'' The trough is formed by two Gaussian laser sheets crossed in a ``V''-shape and propagating along the $x$ axis. These sheets are orthogonally intersected by two parallel vertical sheets propagating along the $y$ axis which complete the three-dimensional trapping potential. All sheets are derived from a single-mode $10\:\text{W}$ laser at $\lambda= 532\:\text{nm}$ and create a repulsive potential for atoms in both the $F=1$ and $F=2$ ground state manifolds. The length of our trough along $x$ is $110\:\mu\text{m}$ and the three-dimensional trapping depth is approximately $10\:\mu\text{K}$. We note that single-photon cooling can be carried out with a variety of dipole trap geometries including an attractive crossed dipole trap \cite{Price} and a fully-enclosed repulsive optical box \cite{Price2}.\par 

We initiate the cooling procedure by adiabatically lowering the magnetic trapping potential. The field is ramped off linearly in time $t_{\textnormal{ramp}}$, which is on the order of one second. During this ramp, the atomic cloud expands and the classical turning point of each atom (in the vertical dimension) approaches the Demon, which is positioned at a fixed distance below the magnetic trap center. To ensure that each atom interacts with the pump beam near its turning point, the adiabaticity condition $\langle \tau_B \rangle / t_{\textnormal{ramp}} \ll 1$ must be satisfied, where $\langle \tau_B \rangle$ is the average oscillation period in the magnetic trap. \par

The pump beam drives a spontaneous Raman transition by exciting the magnetically trapped atoms to the $5^2P_{3/2}(F'=1)$ manifold. From here, the majority of the atoms spontaneously decay to the $F=1$ ground state manifold where they are no longer resonant with the beam. Roughly $16\%$ decay back to the $F=2$ manifold and are subsequently repumped. Because all projections in the $F=1$ manifold $(m_F=-1,0,1)$ couple more weakly to the magnetic field than the initial $|F=2, m_F=2\rangle$ state, they could in principle all be trapped. However, the branching ratios give rise to a final population that is predominantly in the $m_F=0,1$ sublevels.\par

Information about the final distribution of atoms is obtained through absorption imaging. After the cooling sequence, all magnetic and optical fields are switched off and a resonant probe beam propagating along the \textit{z} axis illuminates the atoms for $200\:\mu\text{s}$. The beam is then imaged on a charge-coupled device camera. A variable delay between the field switch-off and the probe illumination allows us to determine the temperature through the time-of-flight method.

\section{Cooling Efficiency}

\begin{figure}
\begin{center}
\includegraphics[width=.7\textwidth]{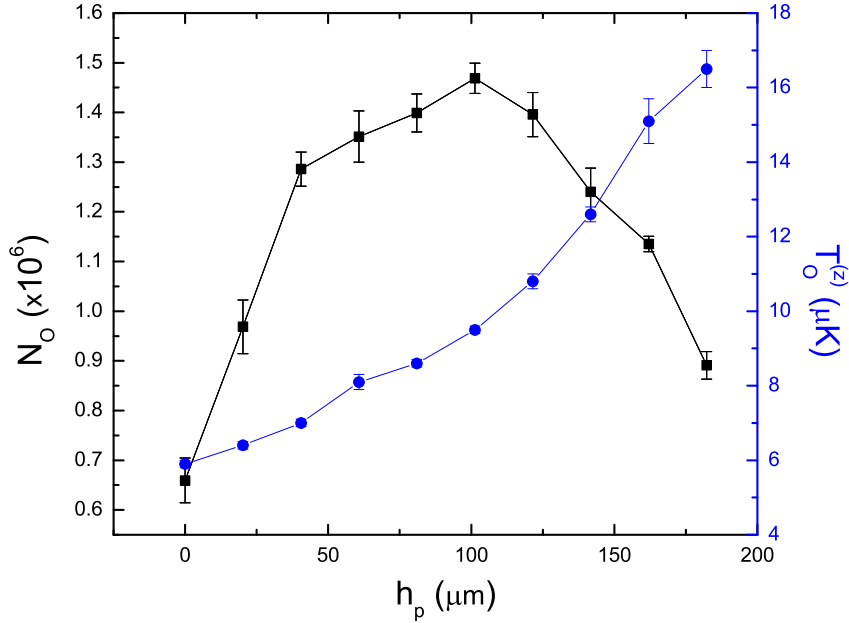}
\end{center}
\caption{\label{fig2}Number (\ding{110}) and temperature (\textcolor{blue}{\ding{108}}) of cooled atoms as a function of $h_p$ (height of the pump beam above the trough vertex).  The positive slope of $T_O^{(z)}$ reflects energy gained by atoms in free fall.  For $h_p > 100\:\mu\text{m}$, the additional energy increases the loss rate from the optical trough.  For $h_p < 100\:\mu\text{m}$ spatial overlap of the pump beam and optical trough beams reduces the excitation probability and hence the capture rate.  The highest phase-space density is achieved at $h_p = 41\:\mu\text{m}$.}
\end{figure}

In order to assess the performance of single-photon cooling, several effects introduced by the geometry of the optical trough should be considered. For example, the height of the pump beam above the trough vertex $h_p$ must be strategically set to optimize cooling. Figure \ref{fig2} shows the effect of $h_p$ on both the vertical temperature $T_O^{(z)}$ and the number $N_O$ of atoms captured in the optical trough. To acquire this data, we image the atoms before they have time to thermalize in the trough. Thus $T_O^{(z)}$ is not an isotropic equilibrium temperature in a thermodynamic sense; rather, it is a measure of the velocity distribution in the vertical dimension. The positive slope of the temperature curve reflects kinetic energy gained by the atoms in free fall. Atoms decaying to the high-field-seeking state ($m_F=1$) gain additional energy from the magnetic field gradient. To obtain the coldest sample possible, one should thus minimize $h_p$ so that the atoms are pumped near the trough vertex. However, the repulsive trough beams overlap the pump beam for small values of $h_p$, lowering the probability of excitation and thereby decreasing the capture number. Maximizing phase-space density ($\rho \propto nT^{-3/2}$ , where $n$ is the atomic density) is accomplished by balancing these two effects.  The point corresponding to the highest phase-space density is located at $h_p = 41\:\mu\text{m}$.\par

It is clear that $T_O^{(z)}$ remains significantly above the recoil temperature (362 nK) even for the smallest values of $h_p$. If we were cooling a one-dimensional ensemble, this extra energy could only be attributed to capturing atoms away from their turning points. However, this effect is negligible as the adiabaticity condition is satisfied: $\langle \tau_B \rangle / t_{\textnormal{ramp}} \approx 5 \times 10^{-3} \ll 1$. Because we are only cooling along the vertical dimension of a three-dimensional magnetic trap, atoms captured in the trough retain energy in the horizontal dimensions. Due to the geometry of the trough, kinetic energy in the $y$ dimension mixes with the $z$ dimension, accounting for the nonvanishing $T_O^{(z)}$.\par

To address the question of transfer efficiency from the magnetic trap to the optical trough, we must consider the phase-space distributions of both. If we model the ensembles in both traps with Maxwell-Boltzmann velocity distributions and Gaussian spatial distributions\cite{footnote1}, the maximum transfer efficiency $\eta$ by loading an optical trap from a magnetic trap through phase-space conserving process may be written

\begin{equation}
\eta \equiv \frac{N_O}{N_B} = \prod_{i=\{x,y,z\}}\frac{\sigma_O^{(i)}}{\sigma_B^{(i)}}\sqrt{\frac{T_O^{(i)}}{T_B^{(i)}}},
\label{eq:1}
\end{equation}

where $N_O$ ($N_B$), $\sigma_O$ ($\sigma_B$), and $T_O$ ($T_B$) are the number, $1/e$ radius, and temperature of the atoms in the optical (magnetic) trap, respectively. The product index $i$ corresponds to orthogonal axes and allows for trap anisotropy, and we assume $(\sigma_O^{(i)},T_O^{(i)})\leq (\sigma_B^{(i)},T_B^{(i)})$.\par

\begin{figure}
\begin{center}
\includegraphics[width=.7\textwidth]{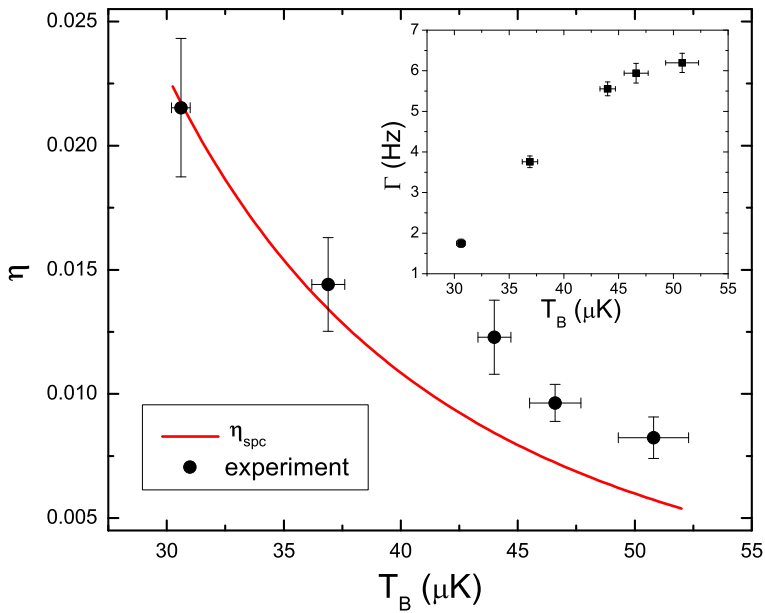}
\end{center}
\caption{\label{fig3}Atom capture efficiency as a function of the initial magnetic trap temperature. The solid line (\textcolor{red}{-}) represents the upper bound capture efficiency given by (\ref{eq:2}). Circles are experimental data. Above $40\:\mu\text{K}$, we measure efficiencies which clearly surpass the predicted limit. We attribute this divergence to an increasing collision rate (inset).}
\end{figure}

In a non-interacting ensemble, single-photon cooling compresses one dimension of the magnetic trap completely in both position and momentum space (neglecting a photon recoil). An upper bound on the efficiency is thus given by (\ref{eq:1}) with the product excluding the compressed vertical dimension:

\begin{equation}
\eta_{spc} = \prod_{i=\{x,y\}}\frac{\sigma_O^{(i)}}{\sigma_B^{(i)}}\sqrt{\frac{T_O^{(i)}}{T_B^{(i)}}}
\propto \left(\sigma_B \sqrt{T_B}\right)^{-2},
\label{eq:2}
\end{equation}

where $T_B = T_B^{(i)}$ indicates a thermalized magnetic trap, and $\sigma_B \equiv \sigma_B^{(x)}=2\sigma_B^{(y)}$ reflects the magnetic trap anisotropy. For a fixed optical trough geometry and depth, $\eta_{spc}$ follows directly from the initial distributions of the magnetic trap. Furthermore, $\eta_{spc}$ may be plotted as a function of $T_B$ by noting that $\sigma_B=\sigma_B(T_B)$ for a thermalized ensemble\cite{footnote2}. In Figure \ref{fig3} we compare experimentally measured efficiencies with the predicted upper bound for several magnetic trap temperatures. \par

The data show fair agreement with (\ref{eq:2}) below $40\:\mu\text{K}$, but there is a trend of increasing efficiency (with respect to the model) for higher temperatures. To explain this trend, we note that our derivation of $\eta_{spc}$ assumes a non-interacting ensemble. The initial trajectories of the ensemble fully determine the dynamics of the cooling process in this case. Only a small fraction of these trajectories, which are represented by (\ref{eq:2}), will become trapped in the trough. In reality, the atoms in the magnetic trap weakly interact through collisions. The single-particle collision rate in the magnetic trap is given by $\Gamma=N^{-1}\int{n(\vec{r})^2 \sigma_s \langle v_r\rangle d\vec{r}}$, where $N$ is the total atom number, $n(\vec{r})$ is the atomic density, $\sigma_s$ is the s-wave scattering cross section, and $\langle v_r \rangle=\sqrt{16 k_B T/\pi m}$ is the mean relative speed in a three-dimensional Boltzmann distribution. The inset in Figure \ref{fig3} shows a monotonically increasing collision rate for increasing magnetic trap temperature. These collisions induce rethermalization of the ensemble, replenishing the trappable trajectories as they are removed from the magnetic trap by the Demon. The end result for a weakly-interacting ensemble is a higher efficiency than predicted by (\ref{eq:2}), which is consistent with the trend in measured efficiencies.\par

Monte-Carlo simulations for an ensemble of non-interacting particles agree with our model and give additional insight into the timescale of single-photon cooling as well as the relevance of collisional interactions. For initial conditions similar to those in Figure \ref{fig3}, simulations indicate an approximately 20\% increase in atom capture by extending $t_{\textnormal{ramp}}$ from $1\:\text{s}$ to $5\:\text{s}$. Extending the ramp time is unadvantageous in practice due to the short lifetime of atoms in the optical trough ($\tau \approx 3\:\text{s}$), which we suspect is due to background gas collisions. In light of the sub-optimal ramp time, it may be inferred from Figure \ref{fig3} that elastic collisions in the magnetic trap play a non-negligible role in the transfer efficiency even at low temperatures. For the measurement at $31\:\mu\text{K}$, collisional gains compensate almost entirely for the sub-optimal ramp time as well as trap losses.

The maximum transfer efficiency we have measured is $2.2(3)\%$. It is clear from (\ref{eq:2}) that the transfer efficiency may be arbitrarily increased by modifying the phase-space overlap of the two traps (e.g. by decreasing the size and temperature of the magnetic trap or increasing the size and depth of the optical trap). One can also use (\ref{eq:2}) to derive a simple expression for the increase in phase-space density of a non-interacting ensemble: $\rho_O/\rho_B=(\sigma_B^{(z)} \sqrt{T_B^{(z)}})/(\sigma_O^{(z)} \sqrt{T_O^{(z)}})$. For a fixed optical trough geometry, this ratio increases with $T_B$ in spite of a corresponding decrease in transfer efficiency.\par

With initial magnetic trap parameters $T_B= 53\:\mu\text{K}$ and $\sigma_B = 515\:\mu\text{m}$, we have transferred $3.3 \times 10^5$ atoms at a temperature of $4.3\:\mu\text{K}$ with $0.3\%$ transfer efficiency. This amounts to a peak phase-space density of $4.9(3) \times 10^{-4}$, which is roughly a $350$-fold increase over the phase-space density of the magnetic trap. The phase-space density is calculated for atoms in the non-magnetic $|F=1, m_F=0\rangle$ state, which accounts for approximately $50\%$ of our final population. This proportion is determined by ejecting the low- and high-field-seeking states from the trough with a large field gradient subsequent to the cooling process.\par

\section{Outlook}
In summary, we have demonstrated a general cooling technique for trapped atoms limited only by the dynamics of the initial trap. We presented an analytical model for the capture efficiency of a non-interacting ensemble and showed that we surpass the limit of the model, likely by means of collisions. Given longer trap lifetimes, these collisions, which facilitate ergodicity, could be exploited to achieve higher phase-space densities and transfer efficiencies. However, we emphasize that although elastic collisions improve the transfer efficiency, they are by no means necessary. The strength of the technique lies in its unrestrictive nature. Because it requires neither a cycling transition nor a scattering cross-section, single-photon cooling can work where other well established methods fail.  

\par

Our technique is particularly promising in light of recent demonstrations with supersonic beams, which have proven the feasibility of producing trapped samples of paramagnetic atoms \cite{Narevicius, Hogan} and molecules \cite{Narevicius2, Bethlem, Meerakker, Sawyer} at tens of millikelvins in a simple room-temperature apparatus. The general nature of single-photon cooling makes it an attractive candidate for cooling and trapping these samples in millikelvin-deep optical traps, the vast majority of which cannot be laser cooled with any other existing technique. Indeed, its implementation has even been proposed for molecules \cite{Narevicius3}, which have been excluded from laser cooling in the past due to complicated energy level structures.\par

We thank E. Narevicius for insightful discussions. We acknowledge support from the R.A. Welch Foundation, the Sid W. Richardson Foundation, and the National Science Foundation. S.T.B. acknowledges support from a National Science Foundation Graduate Research Fellowship.


\section*{References}
\begin{thebibliography}{24}

\bibitem{Szilard} L.~Szilard, Z. Physik \textbf{53}, 840 (1929).
\bibitem{Shannon}  C.~E.~Shannon, Bell System Tech. J. \textbf{27}, 379 (1948).
\bibitem{Jaynes} E.~T.~Jaynes, Phys. Rev. \textbf{106}, 620 (1957).
\bibitem{Bennett} C.~H.~Bennett, IBM J. Res. Dev. \textbf{32}, 16 (1988). 
\bibitem{Scully}  M.~O.~Scully, Phys. Rev. Lett. \textbf{87}, 220601 (2001)
\bibitem{Maxwell}  J.~C.~Maxwell, \textit{Theory of Heat} (Longmans, Green, and Co., London 1875), 4th edition pp. 328 and 329.
\bibitem{Moehl} D.~M\"{o}hl, G.~Petrucci, L.~Thorndahl, and S.~van~der~Meer, Phys. Rep. \textbf{58}, 74 (1980).
\bibitem{Meer} S.~van~der~Meer, Rev. Mod. Phys. \textbf{57}, 689 (1985).
\bibitem{Raizen}  M.~G.~Raizen, A.~M.~Dudarev, Q.~Niu, and N.~J.~Fisch, Phys. Rev. Lett. \textbf{94}, 053003  (2005).
\bibitem{Dudarev} A.~M.~Dudarev, M.~Marder, Q.~Niu, N.~Fisch, and M.~G.~Raizen, Europhys. Lett. \textbf{70}, 761 (2005).
\bibitem{Price} G.~N.~Price, S.~T.~Bannerman, E.~Narevicius, and M.~G.~Raizen, Laser Physics \textbf{17}, 965 (2007).
\bibitem{Ruschhaupt} A.~Ruschhaupt and J.~G.~Muga, Phys. Rev. A \textbf{70}, 061604(R) (2004).
\bibitem{Thorn} J.~J.~Thorn, E.~A.~Schoene, T.~Li, and D.~A.~Steck, Phys. Rev. Lett. \textbf{100}, 240407 (2008).
\bibitem{Price2} G.~N.~Price, S.~T.~Bannerman, K.~Viering, E.~Narevicius, and M.~G.~Raizen, Phys. Rev. Lett. \textbf{100}, 093004 (2008).
\bibitem{Chu} N.~Davidson, H.~J.~Lee, C.~S.~Adams, M.~Kasevich, and S.~Chu, Phys. Rev. Lett \textbf{74}, 1311 (1995).
\bibitem{Ruschhaupt2} A.~Ruschhaupt, J.~G.~Muga, and M.~G.~Raizen, J. Phys. B: At. Mol. Opt. Phys. \textbf{39}, 3833 (2006).
\bibitem{footnote1} {It should be noted that the magnetic and optical trapping potentials are not harmonic, and thus the assumption of Gaussian spatial distributions for our experiment is an approximation. We maintain this approximation to preserve the simplicity and generality of our expression for the transfer efficiency. We estimate a corresponding error of roughly $15\%$, which does not affect the conclusions drawn from comparing the model with experimental data.}
\bibitem{footnote2} {A linear fit of measured radii in this regime yields $\sigma_B=(25.8+5.5T_B\:\mu\text{K}^{-1})\:\mu\text{m}$. The optical trough depths and radii are $(T_O^{(x)},T_O^{(y)}) = (9.5\:\mu\text{K}, 5.2\:\mu\text{K})$ and $(\sigma_O^{(x)}, \sigma_O^{(y)})= (63\:\mu\text{m}, 56\:\mu\text{m})$, respectively.}
\bibitem{Narevicius} E.~Narevicius, A.~Libson, C.~G.~Parthey, I.~Chavez, J.~Narevicius, U.~Even, and M.~G.~Raizen, Phys. Rev. Lett. \textbf{100}, 093003 (2008).
\bibitem{Hogan} S.~D.~Hogan, A.~W.~Wiederkehr, H.~Schmutz and F.~Merkt, Phys. Rev. Lett. \textbf{101}, 143001 (2008)
\bibitem{Narevicius2} E.~Narevicius, A.~Libson, C.~G.~Parthey, I.~Chavez, J.~Narevicius, U.~Even, and M.~G.~Raizen, Phys. Rev. A \textbf{77}, 051401(R) (2008)
\bibitem{Bethlem} H.~L.~Bethlem, G.~Berden, and G.~Meijer, Phys. Rev. Lett. \textbf{83}, 1558 (1999)
\bibitem{Meerakker} S.~Y.~T.~van~de~Meerakker, P.~H.~M.~Smeets, N.~Vanhaecke, R.~T.~Jongma, and G.~Meijer, Phys. Rev. Lett. \textbf{94}, 023004 (2005)
\bibitem{Sawyer} B.~C.~Sawyer, B.~L.~Lev, E.~R.~Hudson, B.~K.~Stuhl, M.~Lara, J.~L.~Bohn, and J.~Ye, Phys. Rev. Lett \textbf{98}, 253002 (2007)
\bibitem{Narevicius3} E.~Narevicius, S.~T.~Bannerman, and M.~G.~Raizen, arXiv:0808.1383 (to appear in New J. Phys.)
\end{thebibliography}
\end{document}